%% file: main.tex
\documentclass[sigconf,nonacm]{acmart}
\usepackage{booktabs}
\usepackage{tabularx}
\usepackage{todonotes}
\usepackage{placeins}
\AtBeginDocument{%
  }

\setcopyright{acmlicensed}
\copyrightyear{2018}
\acmYear{2018}
\acmDOI{XXXXXXX.XXXXXXX}
\acmConference[Conference acronym 'XX]{Make sure to enter the correct
  conference title from your rights confirmation email}{June 03--05,
  2018}{Woodstock, NY}
\acmISBN{978-1-4503-XXXX-X/2018/06}

\newif\ifDEBUG
\DEBUGtrue

\input{misc/typesettings}

\begin{document}

\title{Model-Based Agentic Software Engineering}
\settopmatter{authorsperrow=4}
\author{James C. Davis}
\email{davisjam@purdue.edu}
\affiliation{%
  \institution{Purdue University}
  \city{West Lafayette}
  \state{IN}
  \country{USA}
}

\author{Kelechi G. Kalu}
\email{kalu@purdue.edu}
\affiliation{%
  \institution{Purdue University}
  \city{West Lafayette}
  \state{IN}
  \country{USA}
}

\author{Huiyun Peng}
\email{peng397@purdue.edu}
\affiliation{%
  \institution{Purdue University}
  \city{West Lafayette}
  \state{IN}
  \country{USA}
}

\author{Parth Vinod Patil}
\email{parthvp@amazon.com}
\affiliation{%
  \institution{Amazon Robotics}
  \city{WestBorough}
  \state{MA}
  \country{USA}
}

\renewcommand{\shortauthors}{Davis et al.}

\begin{abstract}
Coding agents increase implementation capacity without automatically making project intent, system structure, or acceptance evidence explicit. 
As implementation becomes abundant relative to engineering judgment, the scarce work shifts toward choosing useful abstractions, producing evidence, and determining which obligations govern acceptance. 
Existing workflows address parts of this gap through larger prompts, repository retrieval, or per-change review, but still require agents and engineers to reconstruct consequential properties.

As an alternative, we present Model-Based Agentic Software Engineering (MAGE).
MAGE is a framework and a theory for building trustworthy autonomy from commodity intelligence.
MAGE addresses a representation problem and an authority problem: it externalizes the smallest purposeful representation needed to answer an engineering question, then gives settled obligations proportionate authority through constraints, sensors, validators, and gates. 
It keeps uncertain intent open and turns recurring reconstruction and judgment into durable engineering structure that later work can inherit.
We developed MAGE from a longitudinal case and refined it through six independently reported industrial accounts. 
Across these sources, MAGE explains how externalized knowledge, bounded action, independent evaluation, and retained human authority can compose into a governed engineering environment, and proposes tests of when that environment turns commodity intelligence into durable engineering progress.
\end{abstract}

\ccsdesc[500]{Software and its engineering~Software development methods}
\ccsdesc[300]{Computing methodologies~Artificial intelligence}

\keywords{agentic software engineering, software models,
  representation engineering, coding agents, engineering governance}


\maketitle


\section{Introduction}

Coding agents are moving software engineering toward a model in which developers
specify goals while agents plan, implement, test, and optimize
software~\cite{jimenez2024swebench, peng2026localcodeoptimizationmultiagent}.
This shift raises a
central question of \emph{assurance at velocity}: \emph{how can we trust systems
that can act with limited human intervention?}
As coding agents increase the rate at which software changes can be produced,
human specification, understanding, and validation can become the limiting
factors in the development process~\cite{monperrus2026endcodereviewcoding, peng-howdo-2026,davis2026cheapcodecostlyjudgment}.

Current approaches primarily improve agents’ access to information through retrieval~\cite{lewis2020neuripsretrievalaugmented}, memory~\cite{hu2026memoryageaiagents}, tools~\cite{yao2023iclrreact}, and harnesses~\cite{Wang_2024, wang2025openhandsopenplatformai}. However, more context does not necessarily make engineering properties explicit. An agent asked whether a change preserves a system boundary may still need to reconstruct that boundary from source code, tests, configuration, and history. A human asked to validate the same change may face the same reconstruction problem. As change volume increases, repeatedly reconstructing and adjudicating these properties becomes a bottleneck for both autonomous reasoning and human oversight.
The challenge is therefore not only to provide more context, but to make consequential engineering properties explicit so they can be reasoned about, evaluated, and governed without repeated reconstruction.

To meet this challenge, we present \emph{Model-Based Agentic Software
Engineering (MAGE)}, a theory of trustworthy agentic software engineering
centered on the governed engineering environment.\footnote{For an expanded
treatment of these topics, see our MAGE book at
\url{https://davisjam.github.io/model-based-agentic-software-engineering/book/mage-book.pdf}.}
MAGE addresses two problems: representation and authority.
Modeling makes consequential engineering knowledge and intent explicit so
humans and agents can reason about larger properties without repeatedly
reconstructing them from code.
Alignment gives important engineering obligations authority through
governance mechanisms (distinguished into constraints, sensors, validators, and gates).
When failures expose missing knowledge or unenforced obligations, MAGE turns
those lessons into durable engineering structure that later work can inherit.
Together, these mechanisms allow greater agent autonomy while reducing the
reconstruction and repeated human judgment needed to govern it.

We developed MAGE in two stages. First, a longitudinal case study of a single project, DocAble, identified underlying mechanisms and their temporal ordering.
This study traced how representations, controls, and recurring judgment evolved within a single system. Second, comparative reconstructions of six independently reported approaches in industry examined whether related structures emerged under different organizational and technical pressures and identified their scope conditions.
Together, these analyses supported the development and refinement of MAGE as a theory of how these structures interact to enable reliable agentic engineering.

This paper contributes:
\begin{itemize}
    \item \emph{MAGE, a theory of trustworthy agentic software engineering}
    that explains how explicit engineering knowledge and authoritative
    obligations compose into a governed engineering environment.

    \item \emph{empirical grounding and refinement of the theory} through a
    longitudinal case study of DocAble and comparative reconstruction of six
    independently reported industrial systems, exposing recurring mechanisms,
    alternative realizations, and scope conditions.

    \item \emph{A falsifiable research agenda} for testing when governed
    engineering environments support human judgment
    and convert agentic capacity into durable engineering progress.
\end{itemize}

\paragraph{Relevance to JAWs.}
We will develop this work into a full TSE manuscript that clarifies the theory and strengthens its methodological and empirical grounding.
We will also elaborate MAGE's relationship to established traditions in software engineering, systems engineering, formal methods, and AI.
JAWs provides an opportunity to obtain early community feedback on a theory.

\section{Background}
\label{sec:foundation}
MAGE builds on governance in software engineering and two established
traditions: engineering approaches for making system knowledge explicit and
enforceable, and AI approaches for representing, retaining, and acting on
state. We first define governance in the operational sense used in this paper,
then summarize the two traditions that MAGE composes.

\subsection{Governance in Software Engineering}
\label{sec:governance-background}

\textit{Governance} describes the problem of making consequential engineering decisions hold as a software system changes: what work may proceed, who may authorize it, what evidence a change must produce, and what conditions admit it. Conventional software engineering provides many mechanisms that serve these purposes, including requirements and architecture, ownership and permissions, review and test policies, continuous-integration gates, release procedures, and incident learning~\cite{winters2020softwareengineeringgoogle}. These mechanisms are distributed across engineering practice rather than organized around a common account of how autonomous work should be governed.

Agentic software engineering makes this problem more acute. Recent work has begun to explore governance mechanisms for autonomous agents, including policies, runtime controls, and human-agent authority boundaries~\cite{kaptein2026runtimegovernance,koch2026governancenorms,ait2025automatedgovernancedslhumanagent}. \emph{Ex ante} mechanisms can encode known obligations before work begins~\cite{koch2026governancenorms}, while \emph{ex post} governance responds when failures expose missing knowledge or unstated obligations; governance conversion makes those lessons durable for later work~\cite{davis2026cheapcodecostlyjudgment}. As implementation outpaces per-change review, the unresolved question is how these functions should be realized in an engineering environment capable of governing autonomous work at scale.

\begin{figure}[t]
  \centering
  \includegraphics[width=\columnwidth,trim=0 180 0 100,clip]{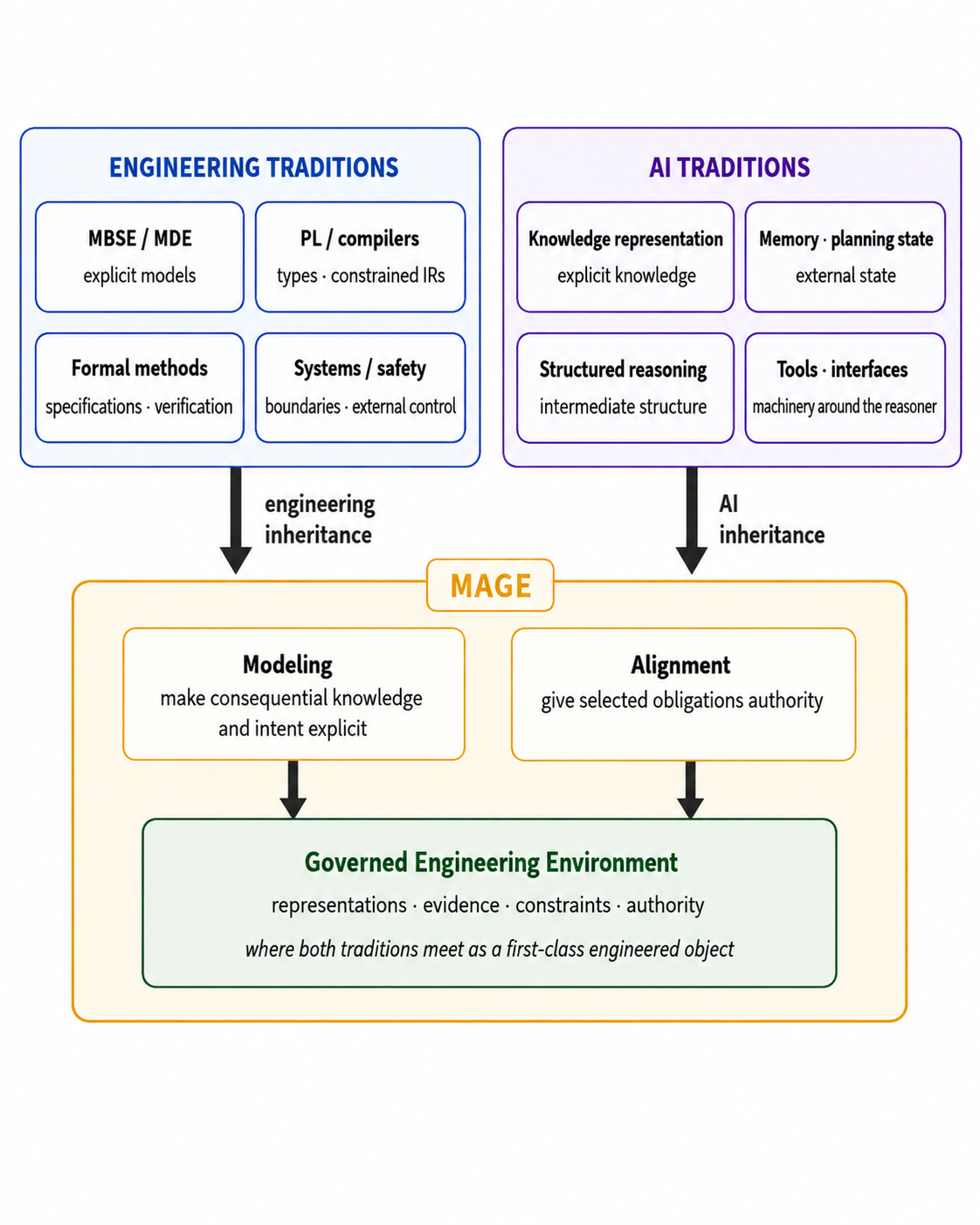}
  \caption{Conceptual foundations of MAGE. The framework combines
  engineering mechanisms for explicit representation, verification,
  and control with AI mechanisms for external state, structured
  reasoning, and tool-mediated action. Modeling and Alignment connect
  these traditions in the governed engineering environment.}
  \Description{Two columns labeled Engineering Traditions and AI
  Traditions feed into MAGE. The engineering column contains MBSE and
  MDE, programming languages and compilers, formal methods, and systems
  and safety. The AI column contains knowledge representation, memory
  and planning state, structured reasoning, and tools and interfaces.
  Within MAGE, Modeling and Alignment both feed a Governed Engineering
  Environment composed of representations, evidence, constraints, and
  authority.}
  \label{fig:mage-traditions}
\end{figure}

\subsection{Engineering Traditions}

Software and systems engineering have long addressed the problem of making
consequential knowledge explicit and actionable. Model-based engineering
externalizes structure and behavior~\cite{schmidt2006mde}; languages and
compilers encode properties through types, interfaces, and intermediate
representations~\cite{pierce2002tapl}; formal methods connect specifications to
verification~\cite{clarke1999modelchecking}; and systems and safety engineering
establish boundaries, hazards, assurance arguments, and independent
controls.
Across these traditions, a representation is a purposeful reduction: a graph,
state machine, contract, or quantitative model preserves what a question needs
while suppressing other detail. It reduces repeated reconstruction and creates
a stable object for reasoning, checking, transformation, and traceability.

\subsection{AI Traditions}

AI provides complementary mechanisms for representing, retaining, and acting
on state. Knowledge representation captures facts and rules~\cite{peng2026sysllmatic};
memory carries information across episodes~\cite{hu2026memoryageaiagents};
planning and structured reasoning organize intermediate
state~\cite{wei2022chainofthought,yao2023iclrreact}; and tools connect reasoning
to repositories, tests, simulators, permissions, and
feedback~\cite{schick2023toolformer}. In coding agents, context and retrieval
shape what is available~\cite{lewis2020neuripsretrievalaugmented}, while
harnesses, workflows, and tools shape what the agent can do and what evidence
it returns~\cite{yang2024sweagent,wang2025openhandsopenplatformai}. The
``Printer'' metaphor captures the resulting division of labor: implementation
becomes abundant, while specifying intent, selecting abstractions, producing
evidence, and determining acceptance remain engineering responsibilities.

\paragraph{The paper's central distinction.}
The operational account above explains what governance must accomplish; the two
traditions supply means for doing so. MAGE combines them in the
\emph{Governed Engineering Environment (GEE)}: an engineered environment
that surrounds autonomous implementation with explicit knowledge and
enforceable obligations. Its novelty is their composition, not any one model or
control. \emph{Modeling} makes consequential intent and system knowledge
explicit in purposeful representations. \emph{Alignment} gives selected
properties operational authority through constraints, sensors, validators, and
gates. The GEE joins the two so engineering knowledge can guide, constrain, and
evaluate autonomous action.

\section{Motivation}
\label{sec:motivation}

Assurance at velocity cannot be inferred from agent capability alone. Coding
agents expand implementation capacity, while the preceding account of
governance shows that explicit knowledge and enforceable obligations reside in
the surrounding engineering environment. An empirical account of outcomes
should therefore distinguish the agentic capability made available from the
engineering environment through which it is used. An adoption indicator marks
access to the former but can leave the latter unspecified.

For example, He \textit{et al.} examined the relationship between Cursor AI
adoption, development velocity, and software quality in open-source
projects~\cite{he2025speedquality}. Their design captures when the tool enters a
project, but not a common software-engineering method or environment through
which it is operated. This distinction is familiar from studies of open-ended
technologies, whose effects depend on how their capabilities are enacted in
practice~\cite{orlikowski2000using,leonardi2011flexible}. Generative AI is
especially open-ended: developers may use the same technology for completion,
debugging, review, testing, optimization, or autonomous implementation, under
different task boundaries, context, oversight, and acceptance criteria.
Projects placed in the same ``adopted'' condition may therefore receive
materially different interventions, limiting what adoption alone can explain.

We believe that the missing empirical object is the \textit{governed engineering environment}: the representations, evidence, constraints, procedures, and authority through which agentic work is reasoned about, constrained, and admitted. Characterizing this environment makes it possible to distinguish access to a general-purpose capability from a reproducible software-engineering intervention. Before its effects can be tested, however, its constructs and mechanisms must be identified. This need motivates the exploratory theory-building design described next.

\section{Methodology}
\label{sec:methodology}

\begin{figure}[t]
\centering
\includegraphics[width=\columnwidth]{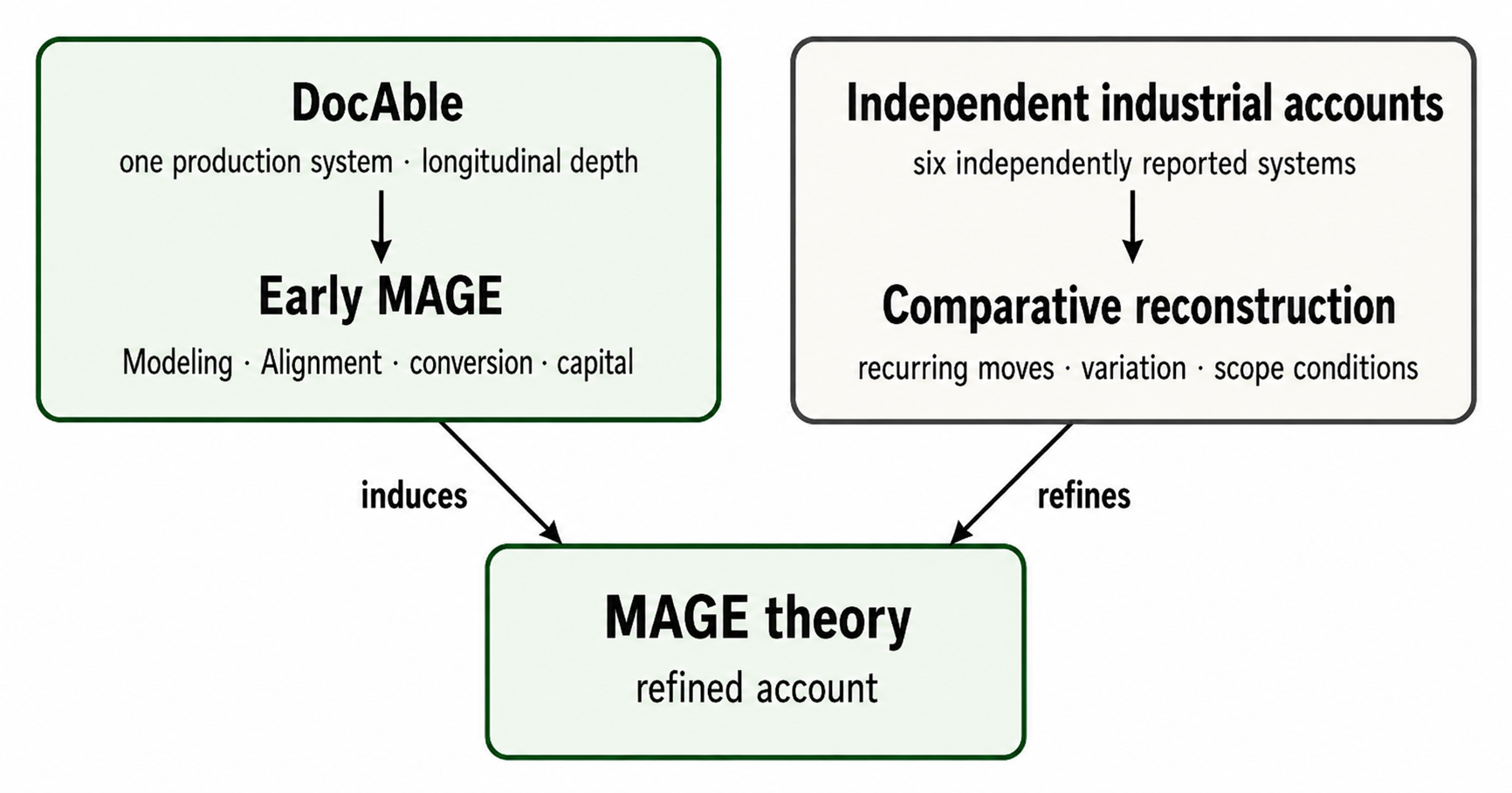}
\caption{Theory-development design. The longitudinal DocAble case induces
initial MAGE constructs, while comparative reconstruction across six
independent industrial accounts refines their recurring mechanisms, variation,
and scope conditions.}
\Description{Two parallel evidence paths converge on MAGE theory. The DocAble
production system provides longitudinal depth and induces early MAGE. Six
independently reported industrial systems support comparative reconstruction,
which refines the theory.}
\label{fig:mage_theory-method}
\end{figure}

This section describes how we developed MAGE (\cref{fig:mage_theory-method}). We use longitudinal evidence from DocAble to identify mechanisms and develop provisional constructs, then compare these constructs with six independently reported industrial systems to examine alternative realizations.

\subsection{Research Design and Evidence}

We use an exploratory, interpretive theory-building design with two sources of
evidence: a longitudinal case of DocAble and comparative reconstructions of six
independently reported industrial accounts. The DocAble case provides the
empirical starting point and temporal record from which the initial mechanisms
and constructs were developed. The industrial accounts provide variation for
examining whether related structures recur under different organizational and
technical conditions.

\paragraph{DocAble}
DocAble is a document-accessibility system developed by the lead author in
response to recent regulatory changes affecting public universities under
Title II of the Americans with Disabilities Act~\cite{doj2024titleii}.
DocAble was developed over approximately 20 weeks of full-time work using
coding agents as the primary implementation workforce. The author exhausted  
3--4 Claude Max 20x accounts during most weeks of the project.
By the end of the observation period, the system comprised approximately
540,000 lines of production code and 1.6 million lines of supporting
governed-environment infrastructure.
Further details of the system and its evolution are available in the MAGE book~\cite{davisMAGE2026}.

Our analysis builds on the case study initially reported in
\cite{davis2026cheapcodecostlyjudgment}, which reconstructed engineering
episodes from contemporaneous field notes and repository history. The present
study does not independently analyze those field notes; it uses that 
case analysis together with the subsequent evolution of the DocAble repository
to develop and refine MAGE.

\paragraph{Comparative sample of industry accounts}
The comparative sample is purposive and comprises first-party accounts from
Cloudflare, Spotify, Shopify, Docker, Siemens, and Zenseact. We selected accounts
that describe autonomous or agentic software work with sufficient architectural
or organizational detail to reconstruct relevant representations, action
boundaries, evaluation mechanisms, and retained human authority. 
These accounts provide variation in engineering pressures, platforms, organizational settings, and approaches to autonomous work. Table~\ref{tab:industrial-reconstructions} previews the
source claims and MAGE interpretations examined in the comparative stage.

\begin{table*}[t]
  \caption{Comparative reconstruction of six independent industrial
  accounts. ``Observed structure'' summarizes selective first-party source
  claims; ``MAGE lens'' is our interpretation. The Siemens source describes prototypes rather than production deployment.}
  \label{tab:industrial-reconstructions}
  \scriptsize
  \renewcommand{\arraystretch}{1.08}
  \begin{tabularx}{\textwidth}{@{}p{0.16\textwidth}p{0.47\textwidth}X@{}}
    \toprule
    \textbf{Account and pressure} & \textbf{Observed structure} & \textbf{MAGE lens} \\
    \midrule
    \textit{Cloudflare}~\cite{cloudflareCodex2026}\newline Institutional standards
       & Structured requirements with stable
      identities, progressive retrieval, and separation of approved
      guidance from enforced obligations. & Modeling institutional
      knowledge plus local Alignment. \\
    \textit{Spotify}~\cite{spotifyHonk2026}\newline Fleet migration  &
      Persistent service and lineage information, fleet targeting,
      build/test feedback, and managed admission. & Platform-level
      knowledge, evaluation, and admission authority. \\
    \textit{Shopify}~\cite{shopifyRiver2026}\newline Organizational knowledge
       & Reproducible repository substrate,
      on-demand skills, durable sessions, searchable work, and useful
      sessions carried into inherited defaults. & Externalized knowledge
      and governance conversion through platform defaults. \\
    \textit{Docker}~\cite{dockerFleet2026}\newline Delegated authority  &
      Bounded roles, tools, and sandboxes; separate producing and reviewing
      agents; human retention of the merge decision. & Alignment through
      action boundaries, independent review, and retained authority. \\
    \textit{Siemens}~\cite{siemensA3E2026}\newline Model-first engineering
       & Task agents operating over persistent
      engineering models, simulation, and expert-defined workflows. & A
      Modeling-first configuration with an explicit evaluation surface. \\
    \textit{Zenseact}~\cite{zenseactZAP2026}\newline Distributed ownership
       & Centrally governed platform with team-owned
      agents and domain knowledge, using progressive disclosure of relevant
      skills and tools. & Federated authority and task-relevant context
      delivery. \\
    \bottomrule
  \end{tabularx}
\end{table*}

\subsection{Analysis and Theory Development}

\paragraph{DocAble theory development.}
The initial analysis of the DocAble record, reported in the Cheap Code
study~\cite{davis2026cheapcodecostlyjudgment}, provides the empirical starting
point for MAGE. That analysis reconstructed salient episodes through their
trigger, interpretation, response, immediate outcome, and effects on later
work, and produced the failure-to-governance process model. Here, that analysis
and the subsequent DocAble record provide an empirical seed rather than a
complete derivation of the theory.

Over the following two months, the author team developed this account
iteratively. We proposed candidate constructs and relationships, applied them
to recurring engineering questions, and revised them when we encountered
contradictions, missing mechanisms, or scope conditions. This process expanded
our earlier emphasis on ex-post
governance~\cite{davis2026cheapcodecostlyjudgment} to include models that make
consequential knowledge available before work begins and the role of
accumulated failure knowledge in Alignment. It yielded provisional constructs
for Modeling, Alignment, governance conversion, and engineering capital.

We also challenged and refined the emerging account through practitioner
conversations, public writing and discussion, and feedback on talks at
organizations including Argonne National Laboratory and IBM Research. These
engagements exposed unclear terms, counterexamples, and alternative
realizations. They informed theory development but are not treated as empirical
observations or independent validation.

\paragraph{Comparative analysis.}
The comparative stage examined recurrence, alternative realizations, and scope
conditions across the industrial accounts. Each account was analyzed using a
common frame: engineering pressure; externalized representation; action
boundaries; evidence and evaluation; admission authority; mechanisms through
which later work inherits prior knowledge; and scope conditions. We kept direct
source claims separate from MAGE interpretations and recorded the analysis in a
case-by-construct matrix linking source material to the refined theory.
Ambiguous and negative observations were retained during comparison.

Across the industrial accounts, we observed recurring structures: knowledge is
externalized, action passes through bounded tools or roles, generation is
separated from evaluation, and consequential authority remains human where the
decision is not adequately mechanized. The accounts arise from different
engineering pressures rather than a shared MAGE adoption program, and several
do not establish model correspondence, longitudinal adaptation, or outcome
measures. The comparison therefore supports claims of recurrence and variation,
not causal effectiveness.

\subsection{Empirical Grounding and Validity}

\paragraph{DocAble grounding.}
The DocAble record provides quantitative grounding for the theory-development
account above. During the 20-week build, 6--8 agents routinely worked in
parallel, sustaining approximately 200 commits per day and 1,000 per week.
This rate exceeded what one engineer could review directly, and quality
degraded as the system grew.
The project responded by shifting increasing
amounts of engineering work into durable models, checks, and controls.

\begin{figure}[t]
  \centering
  \includegraphics[width=\columnwidth]{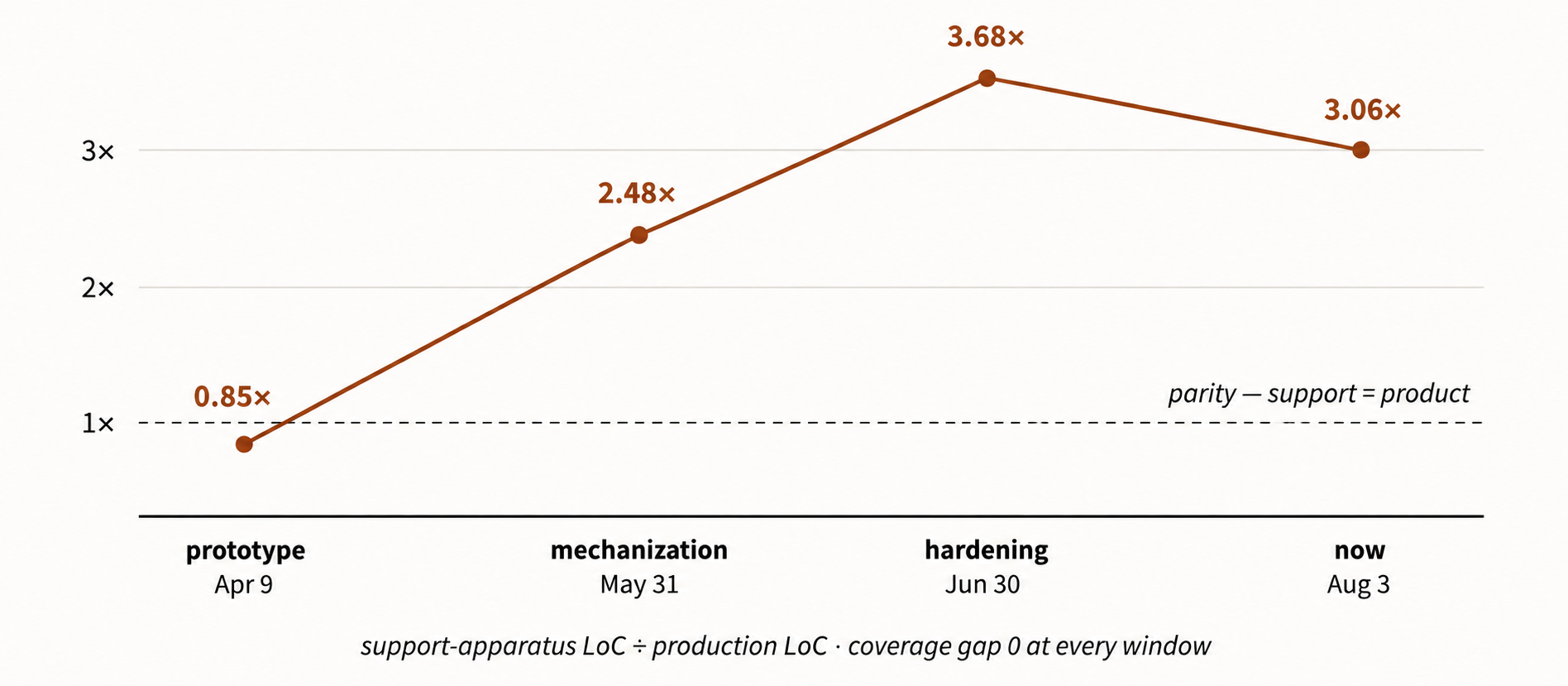}
  \caption{Evolution of DocAble's support ratio across four repository
  snapshots. Support-apparatus source grew from $0.85\times$ production
  source after the prototype to approximately $3\times$ in the mature
  snapshots, peaking at $3.68\times$ during hardening.}
  \Description{A line plot shows DocAble's support-apparatus source divided
  by production source at four snapshots: 0.85 at prototype, 2.48 at
  mechanization, 3.68 during hardening, and 3.06 at the final snapshot. A
  dashed horizontal line at one marks parity between support and production
  source.}
  \label{fig:docable-support}
\end{figure}

This hardening is visible in the composition of the repository. Support
apparatus---tests, models, orchestration, documentation, and governance
tooling---grew from $0.85\times$ production source after the prototype to
approximately $3\times$ in the mature snapshots, peaking at $3.68\times$
during hardening. Figure~\ref{fig:docable-support} shows this trajectory. 

Repository measures show several forms of hardening. Project-specific lint
files grew from 0 to 747 and gate scripts from 0 to 102; 208 commits paired a
fix with a lint intended to catch recurrence. Derived checks caught six
instances of a previously identified class of model--code drift, with no
observed recurrence of that mechanically decidable class across 56 subsequent feature implementations.
Across a nine-stage modeling sequence, the proportion of unmodeled implementation elements decreased from 56\% to 7.89\%.

\paragraph{Validity.}
The industrial
accounts add variation across organizations and engineering settings, with
less visibility into internal failures and evolution. Together, these sources
support theory development, comparative refinement, and analytic
generalization within the stated scope conditions. They do not support
population estimates or causal claims about MAGE's effectiveness.

\section{MAGE Theory}
\label{sec:mage-theory}

MAGE's core claim is that durable engineering progress from agentic work depends not only on agentic capacity, but also on the quality of the engineering environment in which agents operate. This environment make consequential properties answerable and give selected obligations sufficient authority. This section defines MAGE's explanatory model, introduces its core constructs and relationships, and shows how they form feedback loops that distinguish MAGE from a collection of agent tools or governance checks.

\subsection{MAGE's Explanatory Model}

Agentic capacity becomes durable engineering progress only when the environment makes consequential properties answerable and gives selected obligations authority.
Because agentic capacity amplifies its environment, a strong environment turns that capacity into durable throughput, while a weak one produces churn.

MAGE captures these functions through two mechanisms:


\begin{itemize}
\item \textbf{Modeling} externalizes consequential engineering knowledge and intent into explicit, structured models that engineers and agents can reason through, instead of reconstructing meaning from implementation for every question.
\item \textbf{Alignment} gives selected obligations authority through constraints, sensors, validators, and gates that influence agent behavior or determine whether work is accepted.
\end{itemize}

Modeling is the purposeful reduction of system detail around a specific engineering question. A dependency graph can represent architectural relationships, a state machine can specify permitted transitions, a contract can specify interface obligations, and a quantitative model can characterize relevant system behavior. Alignment then connects selected properties to mechanisms that can affect outcomes, such as permission boundaries, tests, admission rules, or deployment gates. Together, Modeling and Alignment determine which properties are both formally representable and operationally consequential within the engineering environment.

\paragraph{Positioning.}
MAGE operates at the level of the engineered environment around autonomous implementation. It complements existing approaches by distinguishing two functions: \emph{Modeling} represents consequential properties, while \emph{Alignment} gives selected properties force through evaluation and control mechanisms.


\paragraph{Scale and assurance strength are independent choices.}
An individual engineer may apply formal verification to a consequential protocol, while a large organization relies primarily on documentation, testing, runtime evidence, and human judgment. 
Broad organizational adoption does not require maximal formalization, just as strong assurance does not require organization-wide adoption. 
MAGE therefore spans a range from lightweight representations of intent and selective checks to formal specifications and trustworthy derivation. 
Across these settings, the underlying engineering problem remains the same: determining what must be represented, what evidence is sufficient, which obligations should have operational authority, and where human judgment remains necessary.

The theoretical contribution lies in this composition of representations and controls over time. Prior knowledge can structure work before execution; representations and controls can guide work during execution; and recurring judgment can be captured as reusable structure for subsequent work. In this way, MAGE extends the earlier middle-range theory~\cite{davis2026cheapcodecostlyjudgment} from reactive governance to an engineering environment that accumulates and reuses judgment.

\subsection{Constructs and Relationships}
The theory centers on a simple mediation claim: \emph{agentic capacity affects realized performance through the governed engineering environment}. The environment does more than constrain work. It determines which properties can be made answerable and which obligations can be given authority. The constructs below define the terms of this mechanism.

\emph{Agentic capacity} is the ability to generate, inspect, and modify software through autonomous or semi-autonomous work. It expands the amount and scope of work that can be attempted, but does not by itself produce durable engineering progress. The \emph{task frontier} is the set of tasks and system scopes that can be undertaken with acceptable risk and human effort. The \emph{governed engineering environment} comprises the representations, evidence, constraints, procedures, and authority through which engineering work is understood, constrained, and admitted.

\emph{Environment quality} has multiple dimensions. \emph{Representation quality} concerns whether the environment exposes the properties relevant to the engineering question. \emph{Obligation coverage} concerns whether important obligations are stated and evaluated during the engineering process. \emph{Coherence} concerns whether representations, implementation, evidence, and controls remain mutually consistent. \emph{Economy} concerns whether the benefits of this structure justify its construction, delivery, and maintenance costs. More artifacts do not necessarily produce a higher-quality environment. The relevant criterion is whether the environment makes the right properties answerable and the right obligations enforceable at acceptable cost.

A central source of poor environment quality is the \emph{semantic gap}: the distance between where an engineering obligation arises and where the environment has sufficient meaning, evidence, and authority to determine whether it is satisfied. When this gap is large, engineers must repeatedly reconstruct context, exercise judgment, or defer decisions to a later boundary where the relevant system state is observable. Modeling reduces the gap by making missing properties explicit. Alignment reduces it by giving obligations operational force through constraints, validators, and gates.

\emph{Realized performance} captures durable throughput and the human attention required to sustain work, rather than implementation volume alone. When attempted work exceeds what the environment can make answerable or enforceable, a mismatch creates \emph{governance pressure}. \emph{Structural diagnosis} identifies the source of this mismatch, such as a local defect, missing representation, weak authority, stale correspondence, an unbridged semantic gap, or an uneconomical mechanism. \emph{Governance adaptation} responds by modifying the environment's models, evidence, procedures, or controls.

Finally, \emph{engineering capital} is durable structure that reduces the cost or uncertainty of future work through reuse. Recurring judgment becomes engineering capital when it is converted into a model, procedure, or mechanism whose expected future value exceeds the cost of constructing and maintaining it.

\subsection{Relationships and Propositions}
These constructs are linked by a directional logic rather than a single fitted causal model. 
Agentic capacity increases the amount of work that can be attempted, while the governed environment mediates whether that work becomes durable performance. 
Modeling contributes representation quality; Alignment contributes authority; and their combination shapes environment quality. 
When governance adaptation converts recurring judgment into economically durable structure, it builds engineering capital and expands the task frontier.

The constructs therefore support four directional propositions.

\textbf{\emph{P1: Environment fit moderates capacity.}} Increasing agentic capacity should produce more durable throughput when the governed environment fits the work, and more churn, escaped failure, or repeated intervention when it does not.

\textbf{\emph{P2: Representation leverage increases with reasoning burden.}} A task-relevant, faithful representation reduces the reconstruction needed to answer an engineering question, with larger gains as system state grows. The claim fails when maintenance and interpretation costs exceed the reconstruction saved.

\textbf{\emph{P3: Authority expands the consequential surface.}} Alignment can make an obligation consequential without a rich system model, while Modeling can make additional system-level obligations legible to Alignment. Their effects are complementary, not interchangeable, and the semantic gap sets how much authority must be deferred or reconstructed rather than enforced in place.

\textbf{\emph{P4: Engineering capital amortizes judgment locally.}} Durable structure should reduce subsequent reconstruction, repeated judgment, rework, or risk on the engineering surfaces that inherit it. Its return should diminish when the structure creates conflicts or costs more to maintain than it saves.

\subsection{Dynamics}

\begin{figure}[t]
\centering
\includegraphics[width=\columnwidth]{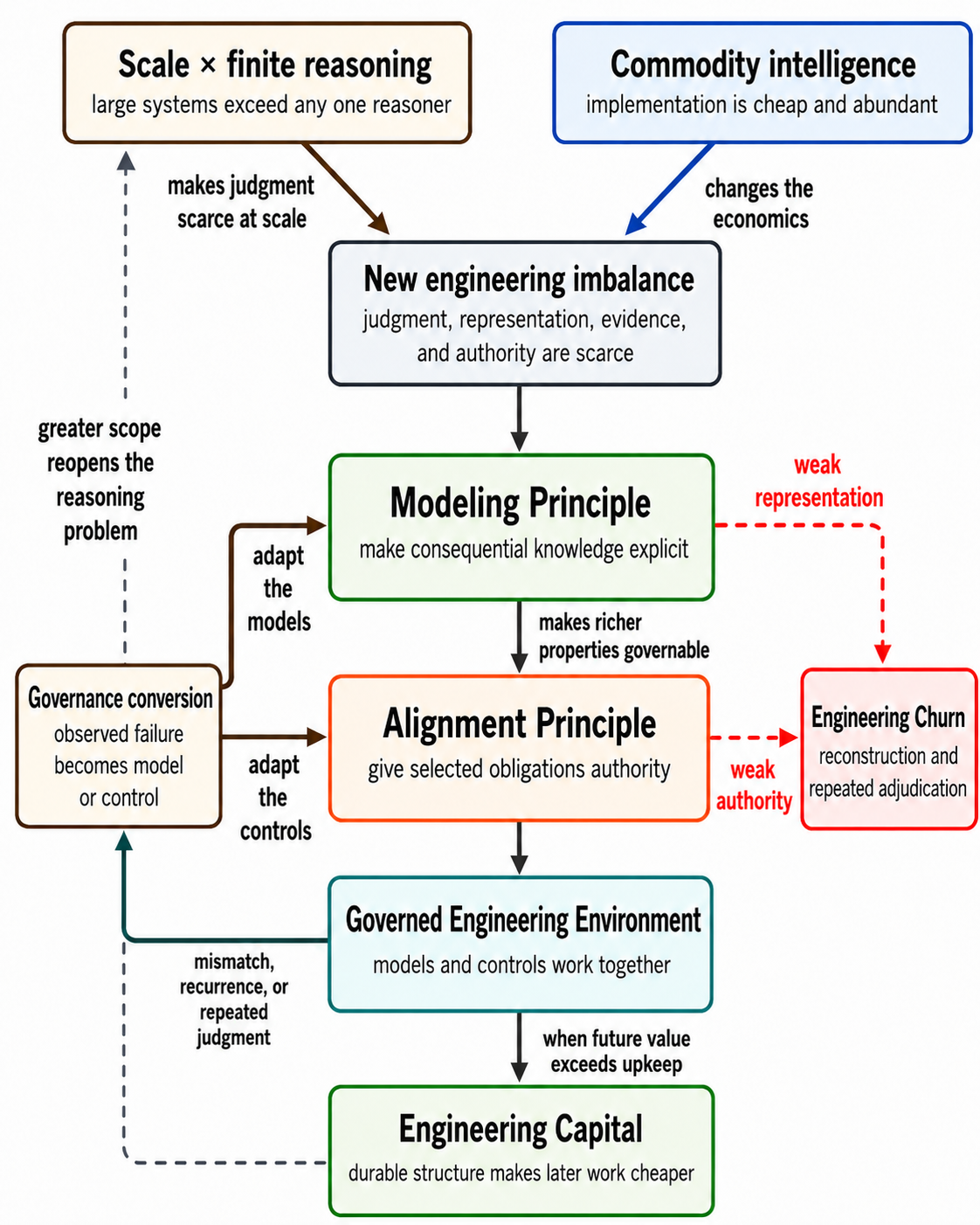}
\caption{Conceptual relationships in MAGE. Scale creates an enduring
reasoning problem, while commodity intelligence makes implementation
abundant relative to engineering judgment. Modeling makes consequential
knowledge explicit; Alignment gives selected obligations authority; and
together they form a governed engineering environment. Governance conversion
turns observed failures into models or controls that later work can inherit.
Weak representation or authority instead produces engineering churn.}

\Description{Scale and finite reasoning and commodity intelligence feed a
new engineering imbalance. A central chain connects the Modeling Principle,
the Alignment Principle, the Governed Engineering Environment, and Engineering
Capital. Governance conversion feeds observed failures back into models and
controls. Dashed red paths connect weak representation and weak authority to
Engineering Churn, while a dashed gray path shows greater scope reopening the
reasoning problem.}
\label{fig:mage-theory-overview}
\end{figure}

Figure~\ref{fig:mage-theory-overview} summarizes the conceptual chain underlying these dynamics. Prior engineering knowledge can bootstrap a governed environment before failures occur. Agentic capacity and attempted work then operate through this environment and shape realized performance. When a task exceeds the environment's ability to make relevant properties answerable or obligations enforceable, the semantic gap becomes visible as governance pressure: the obligation exists, but the environment lacks the representation, evidence, authority, or decision boundary needed to act on it.

Structural diagnosis distinguishes local defects from the broader forms of environment weakness defined above. Governance adaptation responds by changing the models, procedures, evidence, or controls available to future work. In this sense, MAGE treats recurring performance problems not only as implementation errors, but also as potential failures of environment design.

The resulting dynamics contain two distinct feedback paths. The \emph{pressure-and-adaptation} path is balancing: governance pressure prompts changes that reduce the mismatch between attempted work and the environment. The \emph{capability-amplification} path is reinforcing: a stronger environment supports larger tasks or greater concurrency, expanding the task frontier and creating new reasoning and governance demands. Engineering capital can therefore expand the task frontier, but it can also depreciate as the system, obligations, or maintenance costs change.

The model is therefore a directional theory of these relationships. It predicts that durable gains depend on whether the environment makes the relevant properties answerable, gives selected obligations sufficient authority, and does so economically enough for the resulting structure to persist.

\section{Application of MAGE}
\label{sec:application}
In this section, we translate the theory into a practical method. We show how an engineering question determines the appropriate starting point, how Modeling and Alignment apply across project states and scales, and how the same mechanisms can bound autonomous work without requiring a fully formal system model.

\subsection{Engineering Questions $\rightarrow$ Governed Work}

MAGE begins with an engineering question: what to learn, change, preserve, or make true. The engineer externalizes the smallest model needed to answer that question and assigns proportionate authority to settled, consequential obligations. The starting move depends on two conditions: how much implementation already exists and how settled the relevant intent is (Table~\ref{tab:mage-starting-matrix}).

\begin{table}[t]
  \caption{Starting matrix for applying MAGE. Existing system stock
  determines what can be recovered; settled intent determines how much
  representation and authority the work can initially support.}
  \label{tab:mage-starting-matrix}
  \scriptsize
  \renewcommand{\arraystretch}{1.12}
  \begin{tabularx}{\columnwidth}{@{}p{0.20\columnwidth}|XX@{}}
    \toprule
      & \textbf{Intent uncertain} & \textbf{Intent settled} \\
    \midrule
    \textbf{Little implementation} &
      \textit{Explore.} Build enough to learn while preserving known
      constraints. &
      \textit{Model early.} Represent known contracts and consequential
      boundaries before realization. \\
    \textbf{Substantial implementation} &
      \textit{Recover, then explore.} Surface only enough latent structure
      to test competing interpretations. &
      \textit{Recover, reconcile, govern.} Connect representations to the
      realized system before giving them authority. \\
    \bottomrule
  \end{tabularx}
\end{table}

The matrix is diagnostic rather than sequential. Different subsystems may occupy different cells, and a project may move between cells as implementation accumulates or intent stabilizes. 
After the starting move, the same cycle applies: model the reduction that makes the question answerable, align settled obligations with appropriate evidence and authority, perform the work, convert recurring reconstruction or judgment into reusable structure when its future value justifies the upkeep, and reconcile drift. 
The cycle does not require a comprehensive system model; structural, behavioral, ownership, decision, measurement, and provenance views can be connected selectively around the engineering question.

\subsection{Application Across Project States}

\paragraph{Greenfield projects.}
When little implementation exists and intent is uncertain, provisional models and inexpensive implementation support exploration, testing, and refinement. 
As intent stabilizes, established contracts and consequential boundaries can be modeled before realization and used to guide implementation. 
A written hypothesis does not create authority on its own; an obligation becomes authoritative only when its meaning is sufficiently settled and there is adequate evidence to evaluate compliance.

\paragraph{Brownfield projects.}
When substantial implementation exists, the first task is to distinguish what the system does from what the team believes it should do. 
Under uncertain intent, recover only the structure needed to answer the immediate question and test competing interpretations. 
Under settled intent, reconcile the model with the realized system before applying enforcement. 
A bounded migration can use \emph{Audit $\rightarrow$ Drain $\rightarrow$ Promote} to measure the gap, reduce the remaining violations, and then authorize the obligation.

\paragraph{Application across engineering scales.}
The same method can be applied by an individual, team, product, or organization. An individual can model and govern the surface under active change; a team can connect local models to shared architecture, ownership, and operational knowledge; a product can connect representations and obligations across engineering surfaces; and an organization can carry selected policies, assurance obligations, and engineering knowledge across products. The scope and ownership of representations change, but the underlying questions remain: what should be represented, what evidence is sufficient, which obligations warrant authority, and which decisions still require judgment. Stable identities and explicit correspondence allow models at different scales to inform and constrain one another without collapsing into a single organization-wide model~\cite{davisMAGE2026}.

\paragraph{Scale and assurance strength.}
Adoption scope and assurance strength are separate choices. An individual may use formal verification for a consequential protocol, while an organization-wide deployment may rely primarily on design records, tests, runtime evidence, and human admission decisions. MAGE spans a range from lightweight representations of intent and selective checks to formal specifications and trustworthy derivation. The appropriate level depends on consequence, decidability, and maintenance cost.

\subsection{Illustrations of the Method}
\label{sec:evidence}

\paragraph{Why not just prompt?}
At low implementation volume, engineers can compensate for incomplete instructions through review and explanation. 
At agentic velocity, this compensation becomes an assurance bottleneck. 
Prompts, context, and skills can guide work, but they do not by themselves establish correspondence, provide independent acceptance evidence, or make an obligation binding. 
A rule remains guidance unless a constraint, validator, gate, or accountable admission decision gives it operational consequence~\cite{davisMAGE2026}.

\paragraph{Partial automation.}
\label{sec}
Spec-driven development (SDD) shows how parts of this workflow can be proceduralized. 
Kiro externalizes requirements, design, and implementations while retaining human approval at specification boundaries; GitHub Spec Kit provides a related artifact-centered workflow \cite{kiroSpecs2026,githubSpecKit2025}. 
These systems combine forms of Modeling and Alignment, but their specifications require correspondence with the realized system and evaluation.

As part of DocAble, we developed a Claude skill called ``self-governance'' that packages the Modeling--Alignment loop as a procedure an agent can apply to its own work. A recurring Claude Code hook prompts the agent to identify modeling opportunities and missing alignment, allowing it to refine its governed environment. In our use, it reliably proposes local refinements but rarely major changes. Thus, partial automation shifts human effort toward consequential judgment rather than eliminating it.

\section{Research Agenda}
\label{sec:agenda}

We present an initial research agenda for empirically testing selected claims of MAGE.
We focus here on three hypotheses:

\hspace{0.05cm}\textbf{H1---Representation leverage.}
A task-relevant model will reduce reconstruction cost or increase task scale at
matched quality relative to implementation-level context alone; the advantage
will grow with reasoning burden.

\hspace{0.05cm}\textbf{H2---Representation integrity.}
Stale, irrelevant, or poorly chosen models will reduce or reverse H1's benefit,
and correspondence or obligation-coverage failures will predict later churn or
defect escape beyond implementation-level metrics.

\hspace{0.05cm}\textbf{H3---Representation-induced determinization.}
For selected system-level obligations, an explicit model will make a repeatable
predicate feasible where evaluation over raw implementation otherwise requires
semantic reconstruction.

\subsection{Experimental Designs}

\paragraph{Environment-by-agent experiment.}

A first study would compare reasoning engines across otherwise comparable environments with different levels of representation support. Tasks would span local implementation, cross-component changes, behavioral reconstruction, and architectural work. Primary outcomes would include task quality, reconstruction cost, and task scope. The study would test whether representation quality interacts with reasoning burden and agent capability, as predicted by H1.

Our preliminary measurements provide initial evidence in support of this claim. Within DocAble, we conducted a two-week \textit{in situ} experiment in which agents were randomly assigned to receive either explicit instructions for using the models or no such instructions. Bidirectional linkages between code and models ensured that all agents had access to and used the models. Agents receiving explicit instructions nevertheless required fewer tokens, turns, and elapsed time to complete their tasks. The effect was more pronounced for the weaker model (Sonnet).

\paragraph{Representation-integrity intervention.}

A second study would compare synchronized, stale, irrelevant, and absent representations in matched tasks or subsystems. The study would measure correspondence, obligation coverage, reconstruction cost, churn, and defect escape. This design would test H2 and identify the conditions under which representations cease to provide leverage.

\paragraph{Representation-induced determinization study.}

A third study would select system-level obligations for which evaluation currently requires semantic reconstruction. For each obligation, researchers would introduce an explicit model and corresponding predicate, then compare evaluation with and without the model. The primary outcome would be whether the obligation becomes repeatedly evaluable without reconstructing its semantics from the implementation. This study would directly test H3.

\subsection{Measures and Falsification}

Across these studies, evaluation will focus on engineering outcomes. Core measures include reconstruction cost, task quality, task scope, correspondence, obligation coverage, churn, defect escape, and model maintenance cost.

The hypotheses are explicitly falsifiable. H1 is weakened if trustworthy representations do not reduce reconstruction cost or increase task scope. H2 is weakened if representation integrity does not predict downstream failures or if stale and irrelevant representations do not reduce the benefits of H1. H3 is weakened if explicit models do not enable repeatable predicates or if evaluation remains equally dependent on semantic reconstruction.
Together, these studies will establish when representations provide durable engineering leverage, when they become liabilities, and when they can convert system-level knowledge into repeatable evaluation.

\section{Discussion}

\paragraph{Effect size and experimentation}
Software engineering has long sought theories that explain how engineering practices affect software outcomes.
One reason this has been difficult is that, historically, the implementation substrate (the code) was not easily separated from the engineers manipulating it.
Differences among developers, \textit{e.g.}, in expertise and motivation, confounded experiments: human factors often dominate the observable effects of the engineering mechanisms under study. 

MAGE argues that as implementation capacity becomes abundant, the engineered environment becomes increasingly consequential to realized performance.
At the same time, commodity agents make implementation capacity more separable from that environment than human engineering historically allowed.
Hence, compared to many theories of software engineering, MAGE theory has more experimentally separable constructs and is thus more amenable to empirical study.

The same increase in capacity that makes the effect larger also helps expose its mechanisms.
Weak representations, missing constraints, inadequate evidence, and expensive human checkpoints that were tolerable at human implementation rates should increasingly register as bottlenecks as implementation becomes cheap.
Conversely, environments that successfully externalize consequential knowledge and give obligations effective authority should convert more of that capacity into durable progress.
Agentic software engineering may therefore offer something beyond a new phenomenon for software engineering research to explain: it may make longstanding questions about the contribution of engineering structure easier to study.
MAGE proposes one theory of those relationships; its value will depend on whether its predicted mechanisms and directional effects survive controlled variation across agents, environments, tasks, and organizations.


\paragraph{Classical formal methods are one instantiation of MAGE}
MAGE overlaps with formal methods in using explicit representations to support
assurance. 
MAGE characterizes the broader engineering pattern, including heterogeneous representations and enforcement mechanisms. 
Within this pattern, interactive proof assistants~\cite{chlipala2026automaticprogramming} provide a concretization of Modeling--Alignment at one end of the spectrum, favoring formal unification and trustworthy derivation. MAGE's flexibility is deliberate, accommodating the
ambiguity, rapid change, and evolution of ordinary software engineering where grounding everything in one formal logic may not be practical.

\section{Conclusion}

In this paper, we propose Model-Based Agentic Software Engineering (MAGE), a theory of trustworthy agentic software engineering centered on the governed engineering environment. 
MAGE distinguishes two principles: Modeling makes consequential knowledge and intent explicit in task-relevant representations, while Alignment gives selected obligations authority through constraints, sensors, validators, and gates. 
We developed the theory from the longitudinal DocAble case and refined it through comparative reconstructions of six independently reported industrial systems. 

MAGE is intended as a theoretical framework and research agenda for trustworthy agentic software engineering. 
It advances the hypothesis that appropriately engineered representations can reduce the reconstruction required of autonomous agents and enable more durable engineering progress, while alignment mechanisms can translate selected engineering obligations into enforceable action boundaries. 
Testing these claims requires systematic study of how representations, controls, and human decision boundaries shape agent behavior across tasks, projects, and environments.



\balance

\bibliographystyle{ACM-Reference-Format}
\bibliography{references}

\end{document}

%% file: misc/typesettings.tex
\usepackage{algorithm}
\usepackage{xspace}
\usepackage{multirow} 
\usepackage{fancyvrb} 
\usepackage{pifont}

\usepackage{enumitem}
\setlist[itemize]{leftmargin=*,noitemsep,topsep=0pt}
\setlist[enumerate]{leftmargin=*}

\makeatletter
\patchcmd{\@makecaption}
	{\scshape}
	{}
	{}
	{}
\makeatletter
\patchcmd{\@makecaption}
	{\\}
	{.\ }
	{}
	{}
\makeatother

\usepackage{amsthm}

\DeclareMathSymbol{\mlq}{\mathord}{operators}{``}
\DeclareMathSymbol{\mrq}{\mathord}{operators}{`'}

\newif\ifSAVESPACE
\SAVESPACEfalse

\ifSAVESPACE

\else

\fi

\usepackage{soul}
\usepackage{fontawesome}

\ifDEBUG
    \newcommand{\AH}[1]{\todo[color=cyan,inline]{AH:#1}}
    \newcommand{\AM}[1]{\todo[color=red,inline]{Machiry:#1}}
    \newcommand{\JD}[1]{\todo[color=yellow,inline]{JD:#1}}
    \newcommand{\SA}[1]{\todo[color=green,inline]{SA:#1}}
    \newcommand{\PA}[1]{\todo[color=orange,inline]{PA:#1}}
    
    \newcommand{\KR}[1]{\todo[color=yellow,inline]{Kyle:#1}}
    \newcommand{\LS}[1]{\todo[color=green,inline]{LS:#1}}
    \newcommand{\HP}[1]{\todo[color=cyan,inline]{HP:#1}}
    \newcommand{\NJE}[1]{\todo[color=red,inline]{NJE: #1}}
    \newcommand{\GKT}[1]{\todo[color=red,inline]{GKT:#1}}
    
    \newcommand{\RH}[1]{\todo[color=red,inline]{RH:#1}}
    \newcommand{\WJ}[1]{\todo[color=SkyBlue,inline]{Wenxin:#1}} 
    \newcommand{\KC}[1]{\todo[color=orange,inline]{Kelechi Says:#1}}
    \newcommand{\AG}[1]{\todo[color=orange,inline]{AG:#1}}
    \newcommand{\PJ}[1]{\todo[color=lime,inline]{PJ:#1}}
    \newcommand{\AZ}[1]{\todo[color=teal,inline]{Antonio:#1}}
    \newcommand{\PT}[1]{\todo[color=pink,inline]{Parth:#1}}
    
\else
    \newcommand{\AH}[1]{}
    \newcommand{\AM}[1]{}
    \newcommand{\JD}[1]{}
    \newcommand{\SA}[1]{}
    \newcommand{\PA}[1]{}
    \newcommand{\KR}[1]{}
    \newcommand{\LS}[1]{}
    \newcommand{\HP}[1]{}
    \newcommand{\NJE}[1]{}
    \newcommand{\GKT}[1]{}
    \newcommand{\KC}[1]{}
    \newcommand{\RH}[1]{}
    \newcommand{\WJ}[1]{}
    \newcommand{\AG}[1]{}
    \newcommand{\PJ}[1]{}
    \newcommand{\PT}[1]{}
    \newcommand{\AZ}[1]{}
    
\fi

\usepackage{cleveref}
\crefformat{section}{\S#2#1#3}
\crefname{figure}{Figure}{Figures}
\crefname{table}{Table}{Tables}
\crefname{theorem}{Theorem}{Theorems}
\crefname{thm}{Theorem}{Theorems}
\crefname{lemma}{Lemma}{Lemmata}
\crefname{equation}{Eqt.}{Eqts.}
\crefformat{Grammar}{Grammar #1}
\crefname{appendix}{Appendix}{Appendices}
\crefname{listing}{Listing}{Listings}

\usepackage{url}

\makeatletter
\newcommand{\linebreakand}{%
  \end{@IEEEauthorhalign}
  \hfill\mbox{}\par
  \mbox{}\hfill\begin{@IEEEauthorhalign}
}
\makeatother

\usepackage{pifont}

 \usepackage{pifont}